# Observation of transverse spin Nernst magnetoresistance induced by thermal spin current in ferromagnet/non-magnet bilayers

Dong-Jun Kim[1], Chul-Yeon Jeon[1], Jong-Guk Choi[1], Jae Wook Lee[1], Srivathsava Surabhi[2], Jong-Ryul Jeong[2], Kyung-Jin Lee[3,4], and Byong-Guk Park[1, *]

[1] Department of Materials Science and Engineering and KI for Nanocentury, KAIST, Daejeon 34141, Korea

[2] Department of Materials Science and Engineering, Graduate School of Energy Science Technology, Chungnam National University, Daejeon 34134, Korea

[3] Department of Materials Science and Engineering, Korea University, Seoul 02841, Korea

[4] KU-KIST Graduate School of Converging Science and Technology, Korea University, Seoul 02841, Korea

\* Corresponding Author: bgpark@kaist.ac.kr

**Electric generation of spin current via spin Hall effect is of great interest as it allows an efficient manipulation of magnetization in spintronic devices. Theoretically, spin current can be also created by a temperature gradient, which is known as spin Nernst effect. Here, we report spin Nernst effect-induced transverse magnetoresistance in ferromagnet (FM)/non-magnetic heavy metal (HM) bilayers. We observe that the magnitude of transverse magnetoresistance (i.e., planar Nernst signal) in FM/HM bilayers is significantly modified by HM and its thickness. This strong dependence of transverse magnetoresistance on HM evidences the spin Nernst effect in HM; the generation of thermally-induced spin current in HM and its subsequent reflection at the FM/HM interface. Our analysis of transverse magnetoresistance shows that the spin Nernst angles of W and Pt have the opposite sign to their spin Hall angles. Moreover, our estimate implies that the magnitude of the spin Nernst angle would be comparable to that of the spin Hall angle, suggesting an efficient generation of spin current by the spin Nernst effect.**

A central theme of spintronics field is the electrical generation of a spin current as the spin current allows for an efficient magnetization switching and a high speed domain wall motion in magnetic nanostructures[1-6]. In ferromagnet (FM)/non-magnetic heavy metal (HM) bilayers, a longitudinal charge current creates a transverse spin current via spin Hall effect (SHE)[7,8]. The spin current induces spin accumulation at the FM/HM interface, which exerts a torque on the FM and controls the magnetization direction[1,2]. On the other hand, the spin current is partially reflected from the FM/HM interface depending on its spin orientation with respect to the magnetization direction of the FM layer. This reflected spin current is then converted to a charge current via inverse spin Hall effect (ISHE), resulting in the variation of the longitudinal resistance of the FM/HM bilayers, *i.e.*, spin Hall magnetoresistance (SMR)[9-11]. As the SMR originates from the SHE-induced spin current and the ISHE of the reflected spin current, its magnitude depends on the square of the spin Hall angle ($\theta_{SH}$), charge-to-spin conversion efficiency.

Spin current is also generated by a temperature gradient, for instance, the spin (-dependent) Seebeck effect in FM/non-magnetic bilayer structures where thermally-induced spin current is injected from the FM into the non-magnetic layer[12-17]. Theories have predicted that a pure spin current is thermally generated in non-magnetic materials by their spin-orbit coupling effects[18-21], a thermal analog to the SHE, *i.e.*, spin Nernst effect (SNE) (see Fig. 1a). However, there has been no experimental observation yet on the thermally-induced spin current or SNE.

In this work, we demonstrate the SNE by investigating the Hall resistance variation of the FM/HM bilayers under a temperature gradient. Similar to the SMR originating from combined effects of SHE (charge-to-spin conversion) and ISHE (spin-to-charge conversion), the SNE could also cause a resistance variation of the bilayer. This thermally-induced magnetoresistance in a FM/HM bilayer, which can be called spin Nernst magnetoresistance (SNMR), originates from combined effects of two processes: (i) thermally-induced spin current in HM, of which efficiency is described by spin Nernst angle ($\theta_{SN}$), heat-to-spin conversion efficiency, and (ii) subsequent reflection of a spin current at the FM/HM interface and conversion to a charge current via ISHE, of which efficiency is described by $\theta_{SH}$ (see Fig. 1b). As a result, the magnitude of SNMR is determined by the product of $\theta_{SN}$ and $\theta_{SH}$. In analogous to a modification of the planar Hall effect signal (i.e., transverse SMR) by the SHE[11,22], the SNE modifies the planar Nernst effect signal (i.e., transverse SNMR). Therefore, a systematic investigation of transverse SNMR in FM/HM bilayers allows us to identify the SNE, which we have done in this work.

**Results**

**Spin Nernst magnetoresistance model.** The SNMR in a FM/HM bilayer can be described by replacing SHE-induced spin current with thermal spin current, $J_{s,T} = -\theta_{SN}\sigma_{HM}S_{HM}\frac{\partial T}{\partial x}$, in the SMR model[9-11], where $\sigma_{HM}$ and $S_{HM}$ are the electrical conductivity and the Seebeck coefficient of the HM, respectively. The longitudinal ($\Delta V_{xx}$) and transverse ($\Delta V_{xy}$) thermoelectric voltages caused by the longitudinal and transverse SNMRs are respectively expressed as

$$\frac{\Delta V_{xx}}{L_V} = -[S_0 + \Delta S_1 + \Delta S_2(1 - m_y^2)]\frac{\Delta T_x}{L_T}, \quad (1)$$

$$\frac{\Delta V_{xy}}{L_V} = -[\Delta S_2 m_x m_y + \Delta S_3 m_z]\frac{\Delta T_x}{L_T}, \quad (2)$$

where $L_{T(V)}$ is the effective length for temperature gradient (thermal voltage generation), $m_x$, $m_y$, and $m_z$ are the x, y, and z component of the magnetization, respectively, $\Delta T_x$

is the temperature difference along the *x*-axis induced from localized thermal excitation, and

$$\Delta S_1 \equiv -x_{HM}\theta_{SH}\theta_{SN}S_{HM}\frac{2\lambda}{d_{HM}}tanh\left(\frac{d_{HM}}{2\lambda}\right), \quad (3)$$

$$\Delta S_2 \equiv x_{HM}\theta_{SH}\theta_{SN}S_{HM}\text{Re}\frac{\lambda}{d_{HM}}\frac{2\lambda G tanh^2\left(\frac{d_{HM}}{2\lambda}\right)}{\sigma_{HM}+2\lambda G coth\left(\frac{d_{HM}}{\lambda}\right)}, \quad (4)$$

$$\Delta S_3 \equiv -x_{HM}\theta_{SH}\theta_{SN}S_{HM}\text{Im}\frac{\lambda}{d_{HM}}\frac{2\lambda G tanh^2\left(\frac{d_{HM}}{2\lambda}\right)}{\sigma_{HM}+2\lambda G coth\left(\frac{d_{HM}}{\lambda}\right)}, \quad (5)$$

and $S_0$ is the ordinary Seebeck coefficient in the bilayer structure. Here, $\Delta S_1$, $\Delta S_2$, and $\Delta S_3$ are additional Seebeck coefficients induced by SNE, where $d_{HM}$ and $\lambda$ are the thickness and spin diffusion length of the HM, respectively, and $G$ is the spin mixing conductance of the FM/HM interface. Note that SNE in FM layer is ignored and the shunting effect of FM layer is taken into consideration using a geometric factor, $x_{HM}(=\frac{\sigma_{HM}d_{HM}}{\sigma_{HM}d_{HM}+\sigma_{FM}d_{FM}})$, where $\sigma_{FM}$ and $d_{FM}$ are the electrical conductivity and thickness of the FM layer, respectively. The $\Delta V_{xx}$ depends on the magnetization direction relative to the spin orientation (*y*) of SNE-induced spin current, and it is thus proportional to $m_y^2$ while its magnitude is determined by the $\Delta S_2$. As our samples have in-plane magnetization ($m_z \approx 0$), the $\Delta V_{xy}$ ($\propto m_x m_y$) has the same magnitude ($\Delta S_2$) as that of the $\Delta V_{xx}$, so that the investigation of $\Delta V_{xy}$ corresponding to the transverse SNMR (or planar Nernst effect signal) allows us to explore the SNE.

**Angular dependence of transverse spin Nernst magnetoresistance in W/CoFeB structure.** We first examine the transverse SNMR in W(3 nm)/Co$_{32}$Fe$_{48}$B$_{20}$(CoFeB, 2 nm) sample, in which a thermal gradient is generated by a focused laser (55 mW) of ~ 5 μm diameter. Figure 2a schematically illustrates the experiment setup where thermoelectric Hall voltage along the *y*-axis ($V_{xy}$) is measured as a function of in-plane magnetic field angle $\theta$ with respect to the *x*-axis under a temperature gradient. The magnetization is aligned parallel to the applied magnetic field of 100 mT. Depending on the laser position in the sample structure, a vertical ($\Delta T_z$) and/or lateral ($\Delta T_x$) temperature differences in the sample are created accordingly. Upon illumination with a laser spot at the centre of the sample (see Fig. 2a and the top panel of Fig. 2b), generating only $\Delta T_z$ while $\Delta T_x$ cancels out, the thermoelectric signal of W/CoFeB sample shows a clear $cos\theta$ dependence ($\propto m_x$); the largest value (zero) for $\theta = 0$ ($\theta = \pm 90$), where the magnetization is aligned to the *x*-axis (*y*-axis). This reveals that the signal originates from the longitudinal spin Seebeck effect and anomalous Nernst effect[23]. On the other hand, as the laser spot moves toward the edge of the sample, the laser illumination generates non-zero $\Delta T_x$ and as a result, an additional angle-dependent thermoelectric Hall signal appears, which is proportional to $m_x m_y$ or $sin2\theta$. The $sin2\theta$ signal reverses its sign upon the change in the direction of $\Delta T_x$ while the $cos\theta$ signal remains the same sign, which is demonstrated in the middle panel of Fig. 2b where two angle-dependent signals are decomposed. The $sin2\theta$ signal eventually dominates the total signal when the laser spot moves further away, where $\Delta T_z$ induced in the sample is negligible. (see the bottom panel of Fig. 2b), confirming that it originates from $\Delta T_x$. Note that this signal ($\propto m_x m_y$) has the same symmetry as planar Nernst effect (PNE), but the magnitude is noticeably large in the W/CoFeB sample as compared to that in the control sample of a single CoFeB (2 nm) layer (see the decomposed dotted lines in the middle panel of Fig. 2c). As the $\Delta T_x$ in a single CoFeB is comparable to that of W/CoFeB layer (Supplementary Note 1), the large enhancement in PNE indicates that there is a significant contribution from the W layer to the $sin2\theta$ thermoelectric signal, which we attribute to the consequence of transverse SNMR caused by a thermal generation of spin current in W and subsequent reflection of the spin current at the W/CoFeB interface depending on the magnetization direction. Note that the enhanced PNE signal is due to the transverse component of the SNMR as the planar Hall effect in the same W/CoFeB sample is strongly modified by the transverse SMR (Supplementary Note 2). As the transverse SNMR depends on SNE-induced spin current and its conversion into charge voltage via the ISHE, the sign of the transverse SNMR and equivalently the sign of the PNE corresponding to the SNE are determined by the product of the $\theta_{SH}$ and $\theta_{SN}$. For W, it is known as $\theta_{SH} < 0$[11,22,24] and $S_{HM} > 0$[25], thus the positive transverse SNMR for $\Delta T_x > 0$ indicates that a positive $\theta_{SN}$ for W, which is the opposite sign to its $\theta_{SH}$. We note that this sign difference is not impossible because $\theta_{SH}$ is determined by the density of states at the Fermi energy while $\theta_{SN}$ is determined by the energy derivative of density of states[20].

**Material dependence of spin Nernst magnetoresistance.** We also investigate the transverse SNMR for different non-magnetic materials such as Pt and Cu. Note that Pt has a positive $\theta_{SH}$[9,12,24,26], the opposite sign to that of W, while Cu has a negligible $\theta_{SH}$[9,24]. Top panels of Figs. 2d and 2e show that under the central heating, both samples exhibit $cos\theta$ angular dependence as the W/CoFeB or CoFeB sample does (see the top panels of Figs. 2b and 2c). When a sizable $\Delta T_x$ is applied, on the other hand, the $sin2\theta$ thermoelectric signal exhibits a strong material dependence; an opposite sign for the Pt/CoFeB sample and negligibly small for the Cu/CoFeB sample as compared to that of the W/CoFeB sample. As the same thickness of CoFeB is used and a similar $\Delta T_x$ is induced for all samples (Supplementary Note 1), these results again confirm that the $sin2\theta$ thermoelectric signal is dominated by the thermally-induced spin current in HM through its spin-orbit coupling effects.

**Estimation of spin Nernst angle**. We next estimate the heat-to-spin conversion coefficient $\theta_{SN}$ using the HM layer thickness dependence of the transverse SNMR in HM/CoFeB samples. We note that the accuracy of this estimation substantially depends on the accuracy of $\Delta T_x$ and $\Delta T_z$. As it is hard to experimentally determine $\Delta T_x$ and $\Delta T_z$, we estimate the temperature distribution of the sample under the laser illumination by solving the heat transfer module of the COMSOL software (Supplementary Note 1). As a result, we do not argue that our estimation of $\theta_{SN}$ is quantitatively accurate, but we believe that it is still meaningful to estimate $\theta_{SN}$ even approximately.

We performed the same measurement shown in Fig. 2 while varying the laser positions from the centre to the edge of the samples, and then separated the $cos\theta$ and $sin2\theta$ components ($V_\theta$, $V_{2\theta}$). The latter corresponds to the transverse SNMR which is summarized in Figs. 3a and 3b for W/CoFeB and Pt/CoFeB samples, respectively (see Supplementary Note 3 for more details). The $V_{2\theta}$ shows the peak values when the laser is located at the edge of the sample ($x$ ~5 μm), where $\Delta T_x$ is maximized. Figure 3c show the $V_{2\theta}$ of W/CoFeB samples for the edge illumination as a function of W thickness, demonstrating that the $V_{2\theta}$ becomes the largest at 4 nm of W and decreases with a further increase in W thickness. This is the same trend as the W thickness dependence of the SMR in similar W/CoFeB structures[22], indicating that the spin transport in W dominantly governs the transverse SNMR of our samples. A similar thickness dependence of the transverse SNMR is also observed for the Pt/CoFeB samples, which is shown in Fig. 3d. In order to estimate $\theta_{SN}$, we fit the thickness dependence of the transverse SNMR to Equation 2 using material parameters (Table 1) and the calculated $\Delta T_x$ (Supplementary Note 1). Note that the variation of resistivity in W with its thickness has been taken into account (Supplementary Note 4). From the fitting, we obtained $\theta_{SN}$ values of 0.22 ~ 0.41 for W and -0.18 ~ -0.32 for Pt, and $\lambda$ values of (2.0±0.1) nm for W, and (0.8±0.1) nm for Pt. The purple bands in Figs. 3c and 3d indicate error ranges which possibly originates from uncertainties (± 30%) of the literature values of $S_{HM}$, $G$, and $\theta_{SH}$. Note that the Seebeck coefficient of Pt is negative ($S_{HM} < 0$)[25], which is an opposite sign to that of W. This fitting result demonstrates that $\theta_{SN}$ has a comparable magnitude to $\theta_{SH}$ but has an opposite sign to $\theta_{SH}$ for both W and Pt (Table 1). The comparable magnitude between $\theta_{SN}$ and $\theta_{SH}$ implies that the SNE in HM layer can create a spin current as much as the SHE can if a thermal gradient is efficiently generated.

**Discussion**

We demonstrate the transverse SNMR in HM/FM bilayers which signifies an efficient thermal generation of spin current by SNE. Our estimation of the heat-to-spin conversion efficiency $\theta_{SN}$ of W or Pt implies that the magnitude of $\theta_{SN}$ could be comparable to that of the charge-to-spin conversion efficiency $\theta_{SH}$. This suggests that the SNE-induced spin current could create a considerable spin torque to adjacent FM layer, or thermal spin-orbit torques that can manipulate the magnetization direction of the FM as electrical spin-orbit torques do. Moreover, thermal spin-orbit torque can be combined with electrical spin-orbit torque by applying both a charge current and a thermal gradient to bilayers, which allows for the reduction in the critical current for magnetization switching. These results open up an alternative way to generate the spin current and/or to control the magnetization direction in spintronic devices.

*Note added*: We would like to state that while we were preparing the manuscript, we became aware that similar work has been done by other groups, S. Meyer *et al.* Observation of the spin Nernst effect. arXiv:1607.02277 (2016)[27] and P. Sheng *et al.* Signatures of the spin Nernst effect in tungsten. arXiv:1607.06594 (2016)[28].

**Methods**

**Sample preparation.** All samples of W/Co$_{32}$Fe$_{48}$B$_{20}$(CoFeB), Pt/CoFeB, and CoFeB were prepared by magnetron sputtering on thermally oxidized Si substrates with a base pressure of less than 4.0×10$^{-6}$ Pa (3.0×10$^{-8}$ Torr) at room temperature. All samples were covered by MgO (1 nm)/Ta (1 nm) capping layer to prevent oxidation. The bar-shaped structures of 10 μm × 1 mm dimension for spin Nernst magnetoresistance measurement are patterned using photolithography and Ar ion milling. The resistivities are measured to be 320 × 10$^{-8}$ Ω/m for CoFeB, 30 × 10$^{-8}$ Ω/m for Pt, while that of W is 112 × 10$^{-8}$ Ω/m when W is thinner than 4 nm and it gradually decreases with its thickness greater than 4 nm.

**Transverse spin Nernst magnetoresistance measurements.** The thermoelectric Hall voltage along the *y*-axis was measured under the temperature gradients ($\nabla T_x$, $\nabla T_z$) in the sample, which were generated by laser illumination of 55 mW, while rotating a magnetic field of 100 mT in the *x-y* plane. The measurements were repeated at each laser position varying from centre to edge of the sample, which was monitored by its reflectance of the laser. All measurements were carried out at room temperature and each measurement was repeated more than 3 times; data are reproducible.

**Acknowledgements**

This work was supported by the National Research Foundation of Korea (NRF-2015M3D1A1070465, 2014R1A2A1A11051344, and 2016R1A2B4012847)

**Author contributions**

B.-G.P. planned and supervised the study. D.-J.K. and C.-Y.J. fabricated devices. D.-J.K., C.-Y.J., J.-G.C., and J.W.L. performed spin thermoelectric and transport measurement. S.S. and J.-R.J. simulated the temperature distribution. D.-J.K., K.-J.L., and B.-G.P. analysed the results and wrote the manuscript.

**Additional information**

Supplementary information is available in the online version of the paper. Reprints and permissions information is available online at www.nature.com/reprints. Correspondence and requests for materials should be addressed to B.-G.P.

**Competing financial interests**

The authors declare no competing financial interests.

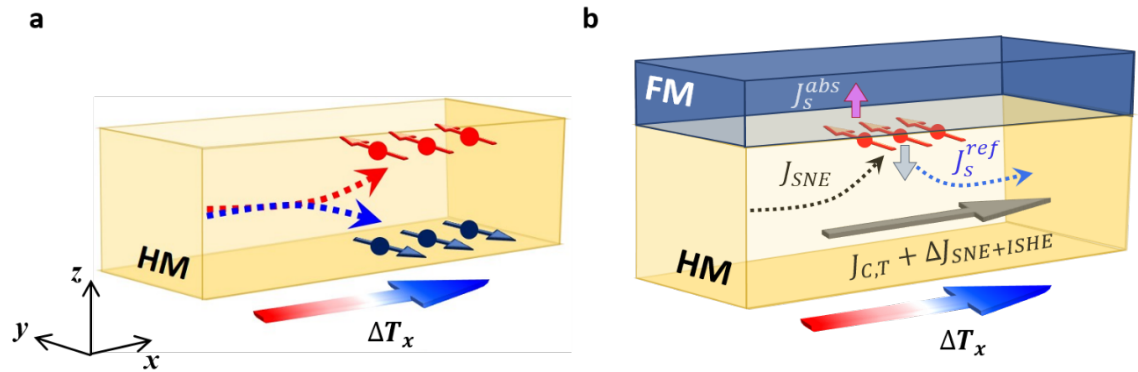

**Figure 1 | Schematics for spin Nernst effect and spin Nernst magnetoresistance.** (**a**) Spin Nernst effect (SNE), where the temperature gradient in *x*-direction generates a spin current in *z*-direction with the spin orientation in *y*-direction. (**b**) Spin Nernst magnetoresistance (SNMR) in FM/HM bilayer where a spin current induced in HM by a temperature gradient in *x*-direction partially reflected at the FM/HM interface depending on its spin orientation with respect to the magnetization direction of the FM layer, resulting in the modification of the longitudinal and Hall resistances of the bilayer.

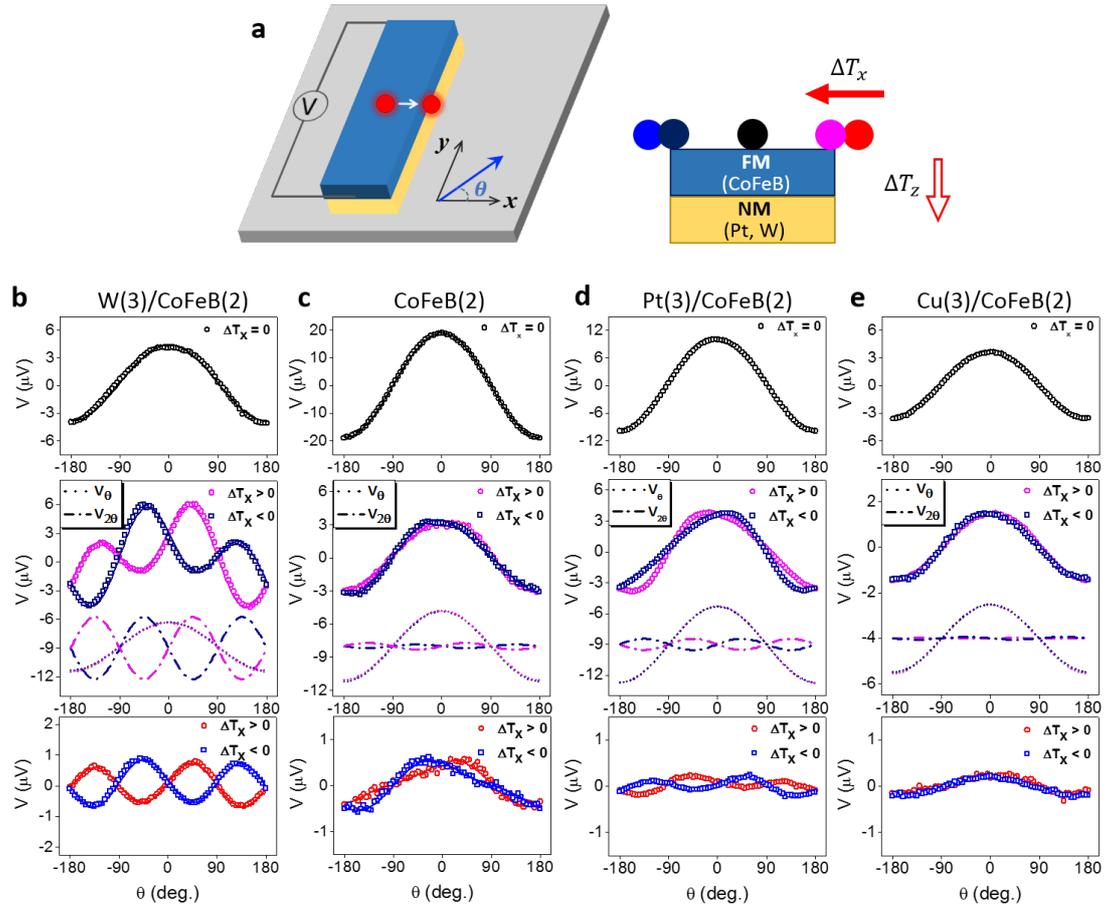

**Figure 2 │ Transverse spin Nernst magnetoresistance originating from SNE in various layer structures.** (**a**) Schematics of measurement under different laser position on bar-shaped structure. The *x-z* plane view indicates the laser positions along the *x* direction. Each color of circle represents the laser position. (**b-e**) Thermoelectric Hall signals for W(3 nm)/CoFeB(2 nm) (**b**), CoFeB(2 nm) (**c**), Pt(3 nm)/CoFeB(2 nm) (**d**), and Cu(3 nm)/CoFeB(2 nm) structures (**e**) for different laser locations, at the centre (*x*~0 μm, top panel), edge (*x*~5 μm, middle panel), and outside of the structure (*x*~10 μm, bottom panel) for each sample. Dotted and dash-dotted lines in the middle panel show the decomposition of two angle-dependent signals of $cos\theta$ and $sin2\theta$. The symbol color denotes the laser position as illustrated in schematics of Fig. 2**a**.

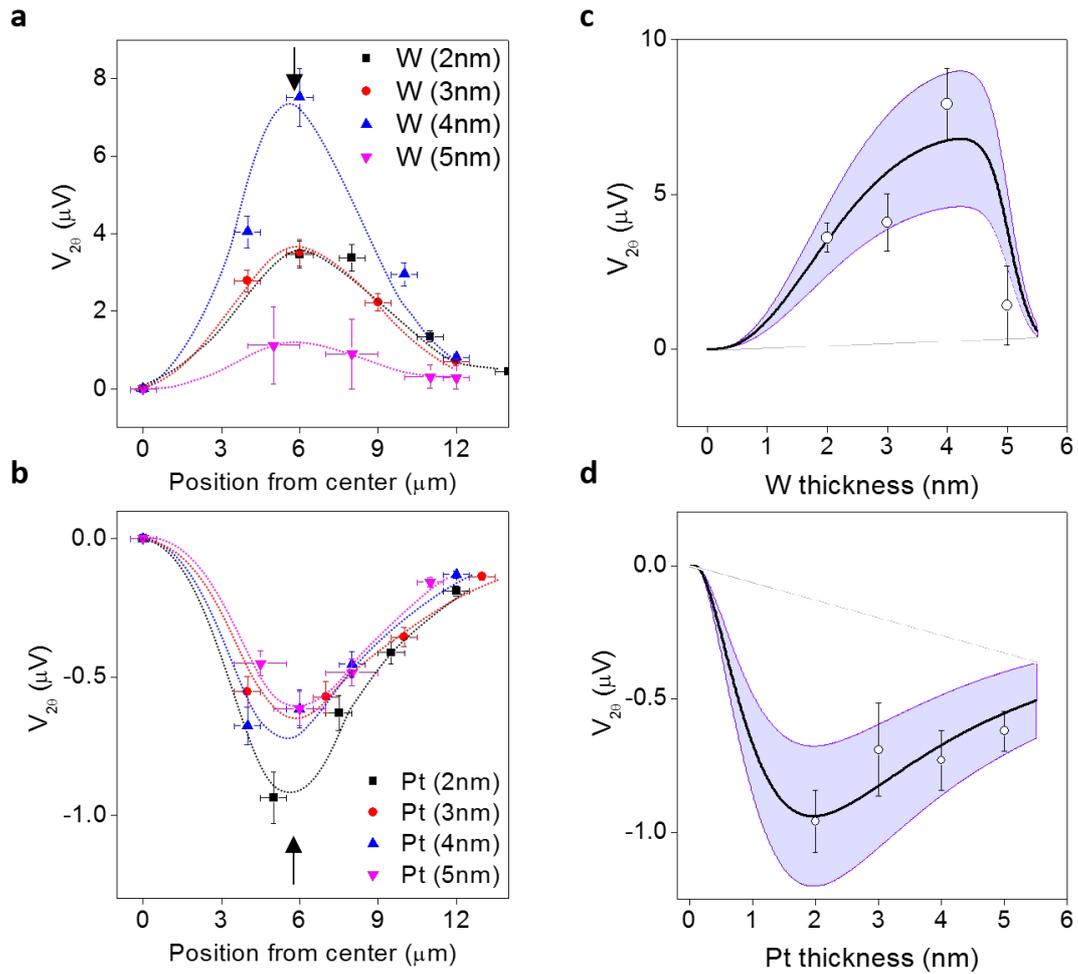

**Figure 3 | Thickness dependence of transverse SNMR in W/CoFeB and Pt/CoFeB structures.** (**a**, **b**) Laser-position dependent thermoelectric signal ($V_{2\theta}$) for W/CoFeB structure (**a**) and Pt/CoFeB structure (**b**) with different HM thicknesses ranging from 2 to 5 nm. Black arrow represents position of edge illumination. (**c**, **d**) HM thickness dependence of the $V_{2\theta}$ for edge illumination for W/CoFeB structure (**c**) and Pt/CoFeB structure (**d**). The white circles represent experimental data and solid lines represent best fitted curves, while purple band indicates error ranges of extracted values, which originated from uncertainties of $S_{HM}$, $G$, and $\theta_{SH}$.

**Table 1. Parameters for analysis of transverse spin Nernst magnetoresistance**

|  | $1/\sigma_{HM}$ [μΩcm] | $1/\sigma_{FM}$ [μΩcm] | $\theta_{SH}$ | $G$ [$\Omega^{-1}m^{-2}$] | $S_{HM}$ [μV/K] | $\lambda$ [nm] | $\theta_{SN}$ |
|---|---|---|---|---|---|---|---|
| W/CoFeB | 35~125 | 320 | -0.21 [22,24] | 0.5~5× $10^{15}$ [11,22] | 10 [25] | 2.0 ±0.1 | 0.22 ~0.41 |
| Pt/CoFeB | 30 | 320 | 0.10 [24,26] | 0.5~5× $10^{15}$ [10,26] | -10 [25] | 0.8 ±0.1 | -0.18 ~-0.32 |

Supplementary Information

# Observation of transverse spin Nernst magnetoresistance induced by thermal spin current in ferromagnet/non-magnet bilayers

Dong-Jun Kim, Chul-Yeon Jeon, Jong-Guk Choi, Jae Wook Lee, Srivathsava Surabhi, Jong-Ryul Jeong, Kyung-Jin Lee, and Byong-Guk Park

## -Contents-

Supplementary Note 1: Temperature distribution generated by laser illumination

Supplementary Note 2: Planar Hall effect for W/CoFeB structure

Supplementary Note 3: Decomposition of thermoelectric Hall signals in HM/CoFeB

Supplementary Note 4: Thickness dependence of the resistivity in W

**Supplementary Note 1. Temperature distribution generated by laser illumination**

We estimated the temperature profiles of CoFeB(2 nm), W(3 nm)/CoFeB(2 nm) and Pt(3 nm)/CoFeB(2 nm) layers on the $SiO_2$(100 nm)/Si(650 μm) substrate using the heat transfer module of the COMSOL software. A monochromatic 660-nm continuous laser beam of 55 mW is considered to have Gaussian power density distribution. The function for the pulse is as follows

$$y(x) = \frac{1}{\sigma\sqrt{2\pi}} e^{\frac{-(x-x_0)^2}{2\sigma^2}}, \tag{S2}$$

where $x_0$ is the centre location of the laser beam, $x$ is the distance from the centre, and $\sigma$ is the standard deviation. The power intensity of the laser decays exponentially from the beam centre in $z$-direction along the thickness of the samples. The Gaussian laser beam has full width half maximum of ~$3\sigma$, equivalent to its experimental diameter of 5μm. The finite element modeling in COMSOL software has been performed using the material parameters obtained under the same experimental conditions as the thermoelectric measurement. The absorption coefficients of W, Pt and CoFeB are $8.66 \times 10^5$ cm$^{-1}$, $7.78 \times 10^5$ cm$^{-1}$ and $8.9 \times 10^5$ cm$^{-1}$, respectively and the total reflectance of CoFeB(2 nm), W(3 nm)/CoFeB(2 nm), and Pt(3 nm)/CoFeB(2 nm) are measured as 0.119, 0.244, and 0.255, respectively. The temperature profiles were calculated utilizing the above parameters for the simulation area of 10 μm × 1 mm, of which size is equivalent to the experimented structure.

The absorbed electromagnetic wave in a thin film deposited on the $SiO_2$/Si substrate is evaluated by the absorption coefficients that is the characteristic feature of particular thickness of respective materials. In an ultra-thin film structure, absorption process could be affected by the multiple reflection and related interference effects, which is considered by solving the Maxwell's electromagnetic wave equations via finite difference time domain (FDTD) optical simulations.

When the laser illuminates on the centre of the sample, a vertical temperature difference $\Delta T_z$ of 33 mK, 40 mK, and 50 mK is generated along the single CoFeB, W/CoFeB, and Pt/CoFeB structures, respectively, while an lateral temperature difference $\Delta T_x$ over the sample structure cancels out, which are illustrated in Supplementary Figs. 1~3. In contrast, when the laser is moved away from the centre of the structure, net $\Delta T_x$ appears. Supplementary Figure 4 shows the lateral temperature differences along the *x*-direction ($\Delta T_x$) which were calculated by integrating temperature over the locally-excited area from $-2\sigma_{sd}$ to $+2\sigma_{sd}$ in the *y*-direction, where the thermal spin current is mostly converted into a transverse voltage. Here, $\sigma_{sd}$ is the standard deviation of the temperature distribution. As a result, similar lateral temperature difference between left and right edge $\Delta T_x$ for both W/CoFeB and Pt/CoFeB samples is obtained to be ~25 K while the $\Delta T_x$ is ~17 K for CoFeB sample. When a laser locates at the edge of the sample, the effective length for temperature gradient and thermal voltage generation, $L_T$ and $L_V$ were defined by $2\sigma_{sd}$ in *x*-direction and $4\sigma_{sd}$ in *y*-direction, respectively.

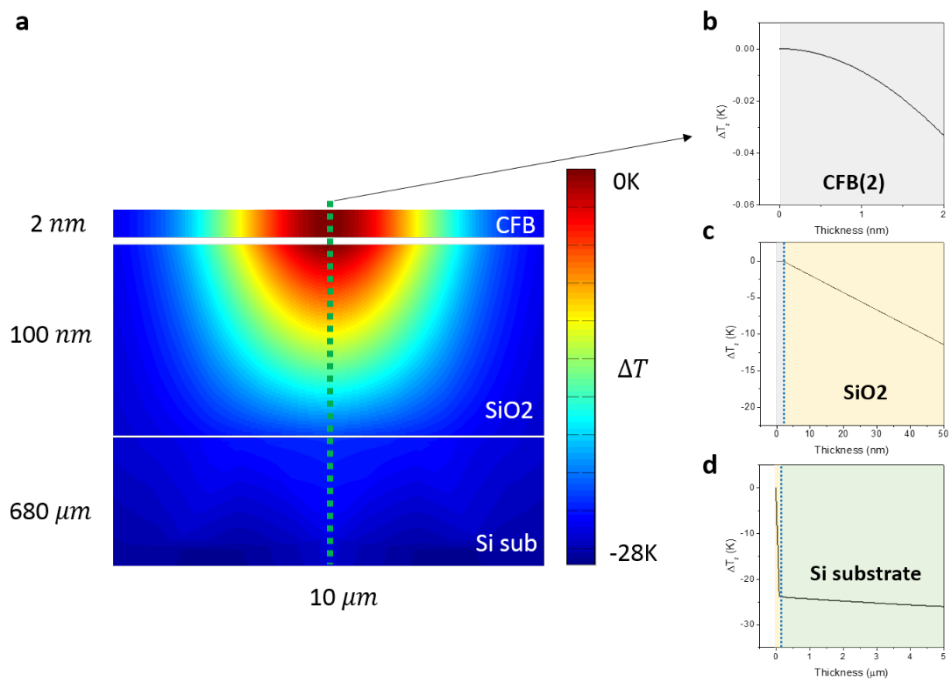

**Supplementary Figure 1 | Temperature profiles in CoFeB (2 nm)/SiO$_2$ (100 nm)/Si substrate.** (**a**) Cross-sectional view of temperature profile calculated under the illumination of a laser spot of 5μm. (**b, c, d**) Temperature profiles along the green-dotted line (vertical temperature distribution at laser centre) within the top CoFeB (2 nm) (**b**), to SiO$_2$ (50 nm) (**c**), to Si substrate (5 μm) (**d**).

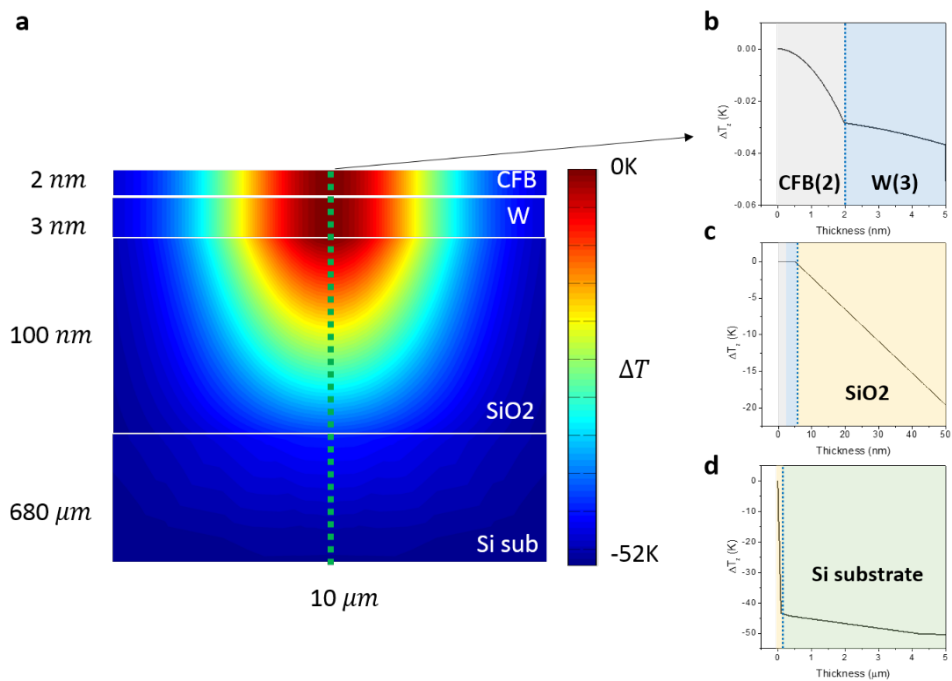

**Supplementary Figure 2 | Temperature profiles in W (3 nm)/CoFeB (2 nm)/SiO$_2$ (100 nm)/Si substrate.** (**a**) Cross-sectional view of temperature profile calculated under the illumination of a laser spot of 5μm. (**b, c, d**) Temperature profiles along the green-dotted line (vertical temperature distribution at laser centre) within the top W (3 nm)/CoFeB (2 nm) (**b**), to SiO$_2$(50 nm) (**c**), to Si substrate (5 μm) (**d**).

.

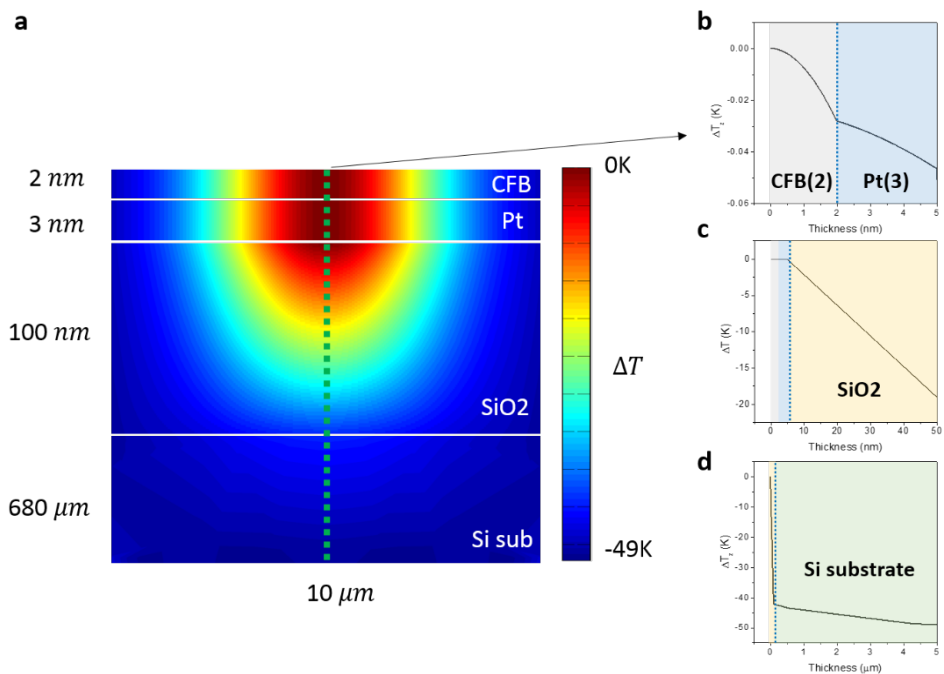

**Supplementary Figure 3 | Temperature profiles in Pt (3 nm)/CoFeB (2 nm)/SiO$_2$ (100 nm)/Si substrate.** (**a**) Cross-sectional view of temperature profile calculated under the illumination of a laser spot of 5μm. (**b, c, d**) Temperature profiles along the green-dotted line (vertical temperature distribution at laser centre) within the top Pt (3 nm)/CoFeB (2 nm) (b), to SiO$_2$ (50 nm) (c), to Si substrate (5 μm) (d).

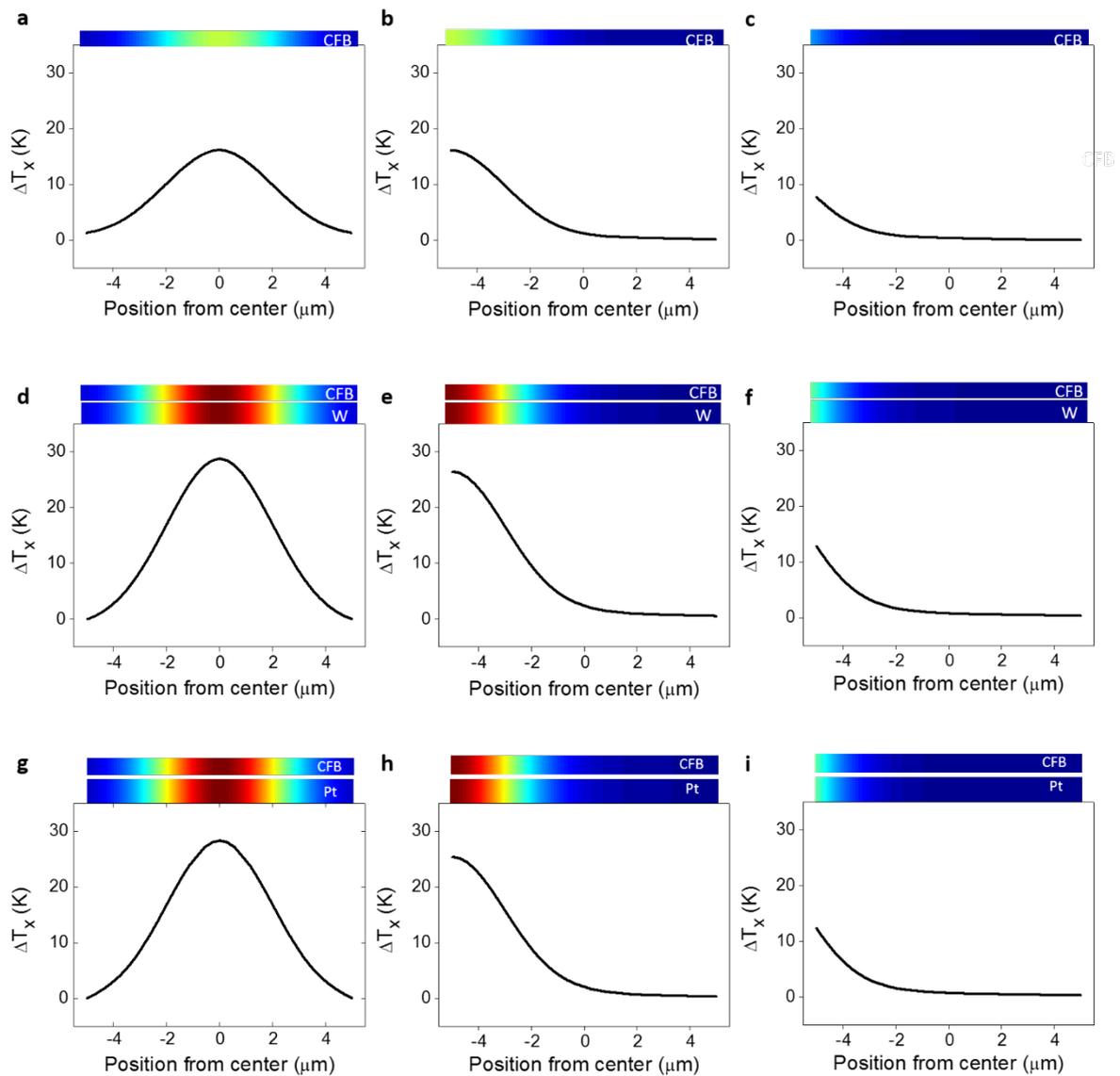

**Supplementary Figure 4 | Effective lateral temperature profiles at different laser locations.** Lateral temperature distribution in (**a, b, c**) CoFeB (2 nm), (**d, e, f**) W (3 nm)/CoFeB (2 nm) and (**g, h, i**) Pt (3 nm)/CoFeB (2 nm) sample at different laser locations. The laser is located on the centre of the sample ($x=0$ μm) (**a, d, g**), on the edge of the sample ($x=5$ μm) (**b, e, h**), and on the just outside of the sample ($x=10$ μm) (**c, f, i**).

**Supplementary Note 2. Planar Hall effect for W/CoFeB structure**

Supplementary Figure 5 shows the planar Hall effect (PHE) in CoFeB (2 nm) and W (3 nm)/CoFeB (2 nm) samples which are the same samples for the transverse spin Nernst magnetoresistance (SNMR) measurement shown in Fig. 2 of the main text. The W/CoFeB sample shows a larger PHE signal than the CoFeB sample by a factor of ~20. In order to study the origin of the enhancement in the PHE, we investigated the angular dependence of the magnetoresistance in the W/CoFeB sample. We measured the longitudinal ($R_{xx}$) and transverse ($R_{xy}$) resistance while rotating the sample on three major planes of the *x-y*, *y-z*, and *z-x* planes under a magnetic field of 9 T. The angle of each plane is denoted as $\alpha$, $\beta$, and $\gamma$, respectively, as indicated in Supplementary Fig. 6. The $\Delta R_{xx}$ represents the spin Hall magnetoresistance (SMR), anisotropic magnetoresistance (AMR), and a sum of the SMR and AMR for the $\beta$-scan, $\gamma$-scan, and $\alpha$-scan, respectively. Supplementary Figure 6 shows that the SMR is much more dominant than the AMR in the W/CoFeB sample. Moreover, the $\Delta R_{xy}$ with $\alpha$ is attributed to the PHE which is the same amount as the $\Delta R_{xx}$, demonstrating the enhancement in the PHE is mostly contributed by the transverse SMR.

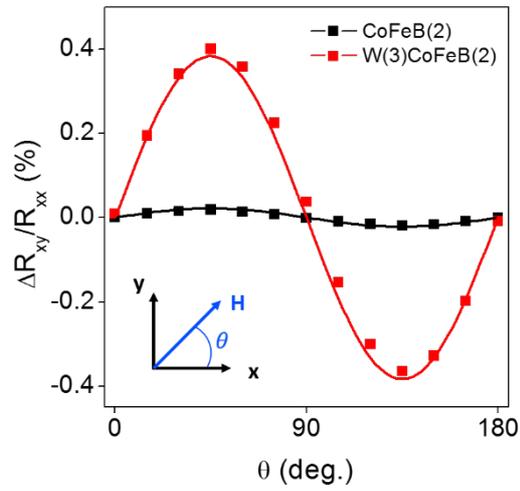

**Supplementary Figure 5 | Planar Hall effect in CoFeB (2 nm) and W (3 nm)/CoFeB (2 nm) samples.** The PHE signal ($\Delta R_{xy}/R_{xx}$) for CoFeB (black) and W/CoFeB (red) is plotted as a function of in-plane angle on *x-y* plane. Black and red curves represent *sin2θ* fitting curves.

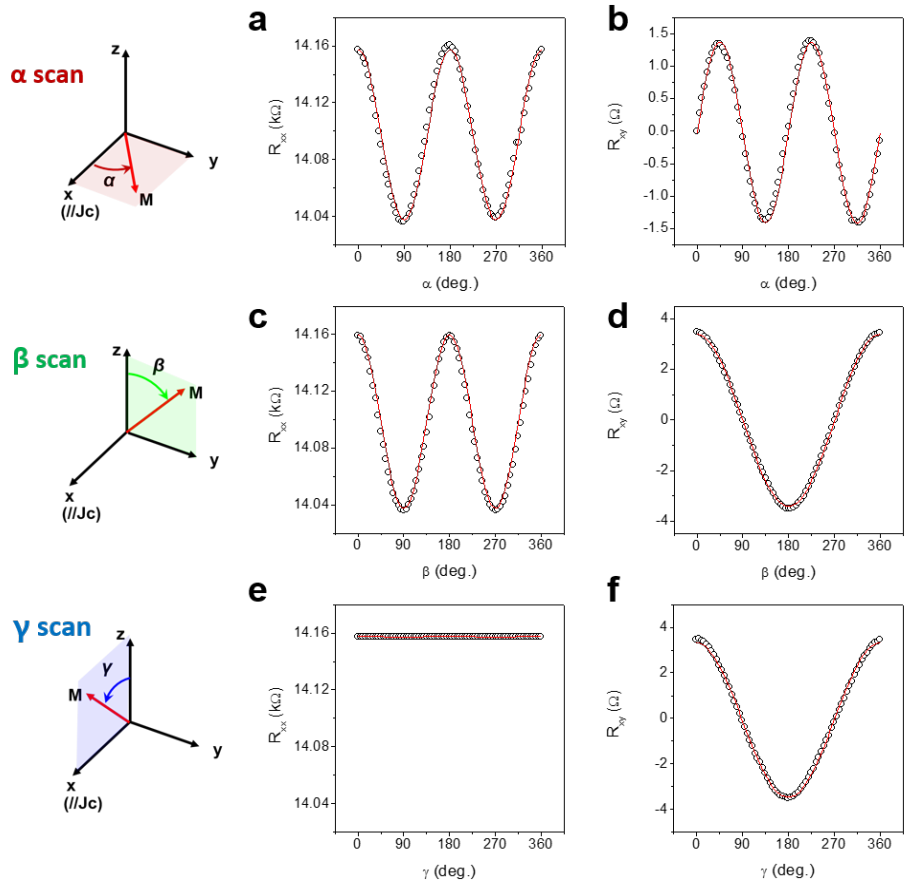

**Supplementary Figure 6 | Angular dependence of the magnetoresistance in W/CoFeB sample.** Angular dependence of the longitudinal resistance $R_{xx}$ (**a, c, e**) and transverse resistance $R_{xy}$ (**b, d, f**) of the W/CoFeB sample for $\alpha$, $\beta$, and $\gamma$ scan, respectively. The measurements were done under a magnetic field of 9 T.

**Supplementary Note 3. Decomposition of thermoelectric Hall signals in HM/CoFeB**

We studied the dependence of the HM thickness on the thermoelectric Hall signal in W($t_W$)/CoFeB and Pt($t_{Pt}$)/CoFeB structures. Supplementary Figure 7 shows the angular dependence of the thermoelectric Hall voltages for each sample at various laser locations. As thermoelectric Hall voltages consist of $cos\theta$ and $sin2\theta$ components ($V_\theta$, $V_{2\theta}$), which are related to $\Delta T_z$ and $\Delta T_x$, respectively, we decomposed them by fitting the measured data with Equation (S1).

$$V = V_\theta \cos(\theta) + V_{2\theta} \sin(2\theta) + C, \qquad \ldots (S1)$$

where $C$ is a constant offset of the thermoelectric voltage. The extracted $V_\theta$ as a function of laser position from the centre of the sample in W/CoFeB and Pt/CoFeB are shown in Supplementary Fig. 8, while the $V_{2\theta}$ components are presented in Figs. 3a and 3b of the main text. The $V_\theta$ in all samples decreases as the laser position moves away from the centre of the sample which is explained by the reduction of the heating area on the structure, or smaller $\nabla T_z$. This indicates that the $V_\theta$ mainly originated from the spin Seebeck effect and anomalous Nernst effect which are proportional to $\nabla T_z$[1,2].

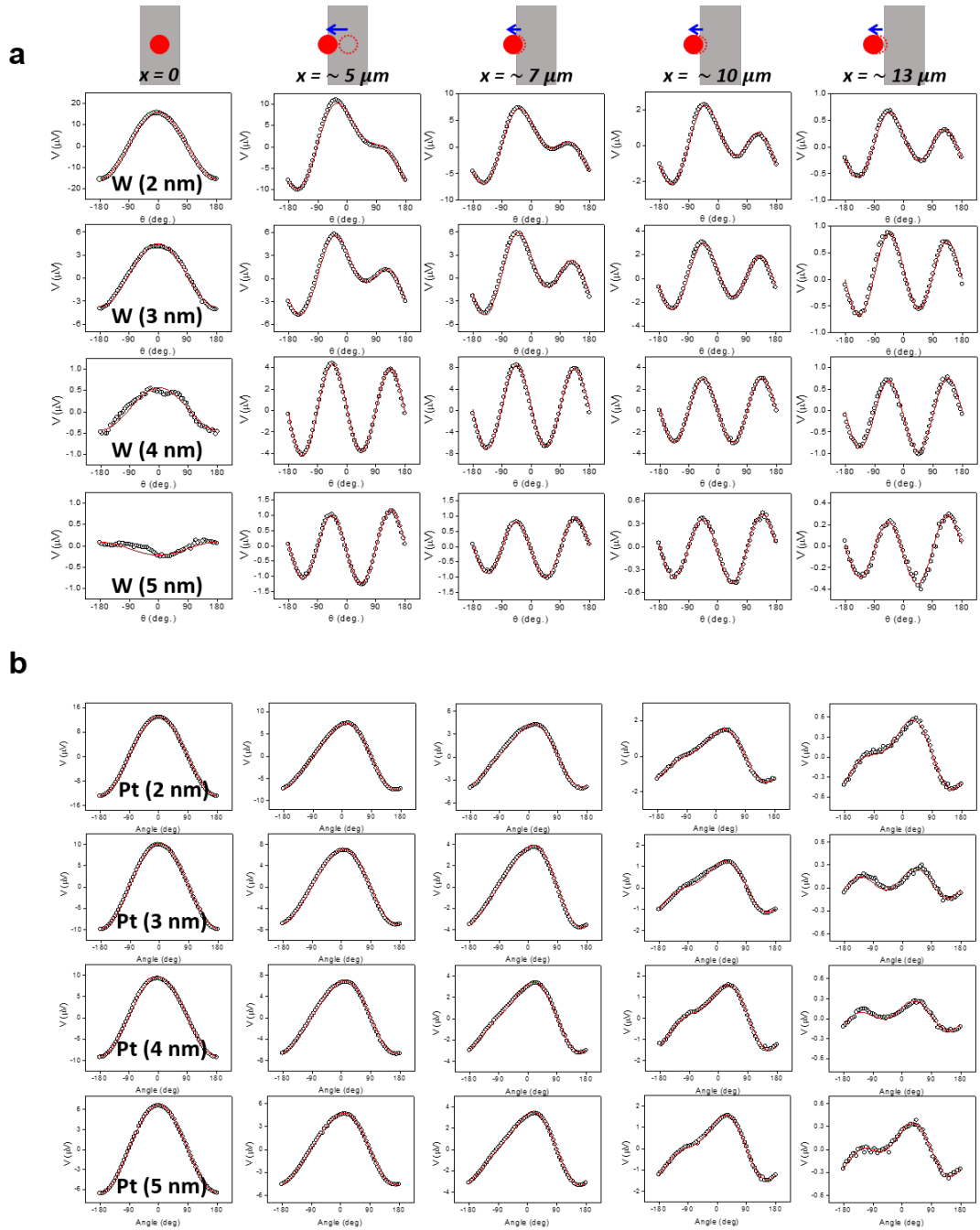

**Supplementary Figure 7 | Thermoelectric Hall signals in W/CoFeB and Pt/CoFeB samples.** (**a, b**) Thermoelectric Hall signals for W($t_W$)/CoFeB(2 nm) with $t_W$=2~5 nm (**a**) and Pt($t_{Pt}$)/CoFeB(2 nm) with $t_{Pt}$=2~5 nm (**b**). The red lines are fitting curves using Equation S1. Top schematics illustrate the laser position of each measurement.

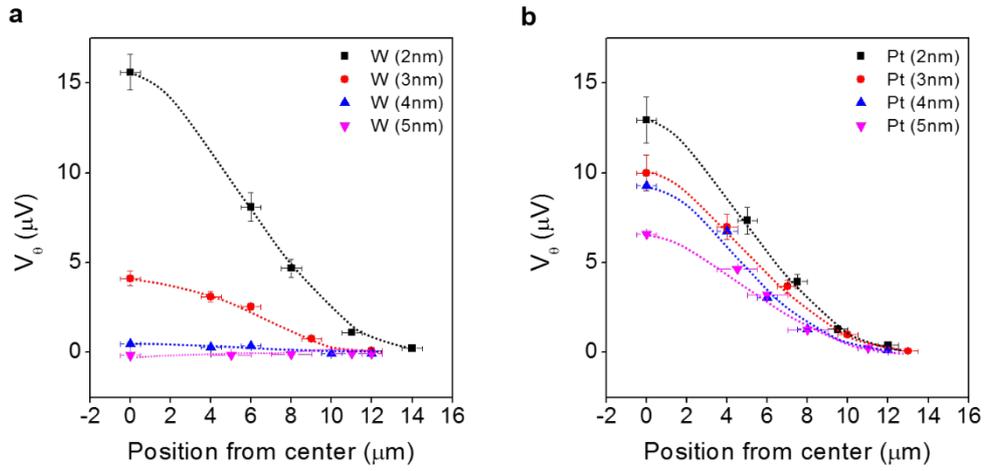

**Supplementary Figure 8 | Laser position dependence of the thermoelectric Hall signals. (a, b)** The extracted $V_\theta$ as a function of distance of the laser position from the centre of the sample for W/CoFeB (**a**) and Pt/CoFeB structures (**b**).

**Supplementary Note 4. Thickness dependence of the resistivity in W**

The resistivity of W varies with its thickness. We measured the resistance of the W/CoFeB samples as a function of W thickness, demonstrating that the resistivity starts to decrease for the W thickness larger than 4 nm (see Supplementary Fig. 9a). This is attributed to the phase change in W; from β-W to α-W as the thickness increases[3]. The variation of the resistivity alters the geometric factor $x_{HM}$ as well (see Supplementary Fig. 9b). The resistance and $x_{HM}$ variations in W/CoFeB samples were taken into account when the spin Nernst angle was estimated in the main text.

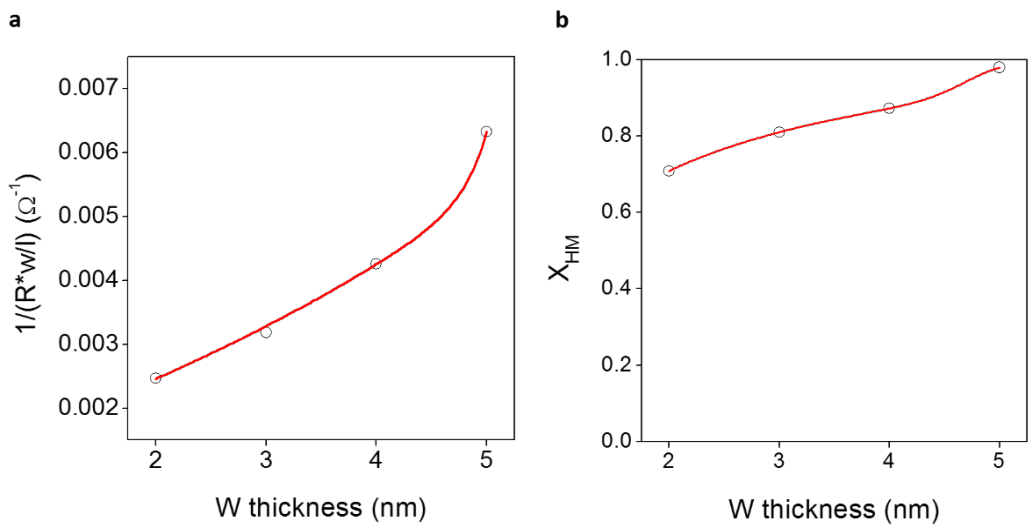

**Supplementary Figure 9 | W thickness dependence of resistivity and geometric factor.** (**a**) Resistivity in W electrode as a function of its thickness. (**b**) $x_{HM}$ variation as a function of W thickness.